\documentclass[11pt,twoside]{article}
\usepackage{asp2010}

\resetcounters

\begin{document}

\title{Circuit Design: An inquiry lab activity at Maui Community College}
\author{Katie Morzinski$^{1,2}$, Oscar Azucena$^{1,3}$, Cooper Downs$^4$, Tela Favaloro$^3$, Jung Park$^5$, and Vivian U$^{4,6}$
\affil{$^1$Center~for~Adaptive~Optics, University~of~California, Santa~Cruz, CA 95064}
\affil{$^2$Astronomy~Dept., University~of~California, Santa~Cruz, CA 95064}
\affil{$^3$Electrical~Engineering~Dept., University~of~California, Santa~Cruz, CA 95064}
\affil{$^4$Institute~for~Astronomy, University~of~Hawai`i~at~Manoa, 2680 Woodlawn~Dr., Honolulu, HI 96822}
\affil{$^5$Electronic~and~Computer~Engineering~Technology, University~of~Hawai`i, Maui~College, 310 Ka`ahumanu Ave., Kahului, HI 96732}
\affil{$^6$Harvard-Smithsonian~Center~for~Astrophysics, Mail~Stop~66, SAO, 60~Garden~St., Cambridge, MA 02138}}

\begin{abstract}
We present an inquiry lab activity on Circuit Design that was conducted in Fall 2009 with first-year community college students majoring in Electrical Engineering Technology. This inquiry emphasized the use of engineering process skills, including circuit assembly and problem solving, while learning technical content. Content goals of the inquiry emphasized understanding voltage dividers (Kirchoff's voltage law) and analysis and optimization of resistive networks (Th\'evenin equivalence). We assumed prior exposure to series and parallel circuits and Ohm's law (the relationship between voltage, current, and resistance) and designed the inquiry to develop these skills. The inquiry utilized selection of engineering challenges on a specific circuit (the Wheatstone Bridge) to realize these learning goals. Students generated questions and observations during the starters, which were categorized into four engineering challenges or design goals. The students formed teams and chose one challenge to focus on during the inquiry. We created a rubric for summative assessment which helped to clarify and solidify project goals while designing the inquiry and aided in formative assessment during the activity. After describing implementation, we compare and contrast engineering-oriented inquiry design as opposed to activities geared toward science learning.
\end{abstract}

\section{Introduction}
Maui, the second-most densely populated island in the state of Hawai`i, hosts a suite of research telescopes on the 10,000-foot summit of Haleakal\=a operated by the University of Hawai`i's Institute for Astronomy (IfA) and the U.S.\ Air Force. Science, technology, engineering, and mathematics (STEM) employers on the island include the observatories at Haleakal\=a as well as opportunities such as the Maui High Performance Computing Center at the Maui Research and Technology Center. Working to prepare Maui residents for Maui-based STEM careers are the Akamai Workforce Initiative (AWI) and the Institute for Scientist and Engineer Educators (ISEE). ISEE has grown out of the former Center for Adaptive Optics (CfAO) Professional Development Program (PDP, \citeauthor{PDPdesc}\ \citeyear{PDPdesc}) and other education programs. Among the sponsors of these programs are the National Science Foundation, the University of Hawai`i, and the Air Force.

Undergraduate students on the compact island are served by
the University of Hawai`i, Maui College (formerly Maui Community College)
in the central town of Kahului. At the time of teaching, the institution was called Maui Community College (MCC) and students could obtain an Associate's degree in Electronics and Computer Engineering Technology. As part of the Akamai Workforce Initiative and in partnership with ISEE, CfAO, and IfA,
the process has begun to convert MCC to a four-year institution called the
University of Hawai`i, Maui College.
UH-Maui will offer a Bachelor's degree in Applied Science in Engineering Technology.
Toward this end, PDP and ISEE participants have been designing inquiry lab activities for
UH-Maui to grow its curriculum in electro-optics technology.

An inquiry lab teaches process skills and content knowledge in a particular STEM field
by engaging students in learner-directed activities
that mirror authentic science and engineering \citep{what_is_inquiry, elements_of_inquiry}.
Targeted facilitation is used to guide students toward the learning goals.
This paper describes an inquiry on circuit design prototyped in Fall 2009 at MCC.

\section{Activity Description}

\subsection{Overview}
This activity fits into a formal course introducing electric circuits to college students majoring in Electrical Engineering Technology,
and assumes some prior exposure to Ohm's law (voltage is equal to current times resistance) and to series and parallel circuits.
A particular circuit (the Wheatstone Bridge) is used to study the content goals of voltage dividers and analysis of resistive networks.
In the activity ``Starters'', the Wheatstone Bridge circuit is introduced.
This circuit features a voltage reading across the bridge that is highly sensitive to small changes in resistance.
Design goals are presented as engineering challenges, and student teams choose one to address.
In the Focused Investigation, teams work toward meeting their design goal
by building, measuring, and analyzing their own variation on the Wheatstone Bridge circuit.
Materials required are breadboards, wires and connectors, resistors,
rheostats or potentiometers, thermistors, multimeters, and power supplies.
The duration of the lab is two 105-minute class periods.
Table~\ref{tab:timeline} shows the activity timetable.

\begin{table}[htbp]
  \caption{
  Timeline for circuit design inquiry.
  }
  \smallskip
  \centering
  \begin{tabular}{lclc}
    \hline
    \multicolumn{2}{l}{\textbf{Day 1}} & \multicolumn{2}{l}{\textbf{Day 2}} \\
    \hline
    Introduction & 15 min. &     Thinking tool & 10 min.\\
    Starters & 20 min. &    Focused investigation & 40 min.\\
    Break & 10 min. &    Break & 10 min.\\
    \multicolumn{2}{l}{(Facilitators sort questions)} &     Poster preparation & 15 min.\\
    Choosing a design goal & 10 min. &   Sharing & 15 min.\\
    Focused investigation & 50 min. &  Synthesis & 15 min.\\
    \hline
     & \textit{Total Time} & \textit{3.5 hrs.} & \\
    \hline
  \end{tabular}
  \label{tab:timeline}
\end{table}

\subsection{Venue}
This inquiry lab activity was designed by Oscar Azucena (Design-Team Leader),
Cooper Downs, Tela Favoloro, Katie Morzinski,
Jung Park, and Vivian U as a new activity during the 2009 PDP.
It was taught at MCC in Professor Mark Hoffman's class \textit{Electronics 101: Introduction to Electronics Technology}
on Tuesday and Thursday, the 27th and 29th of October 2009.
There were twelve primarily first-year undergraduate students, interested in majoring in Electrical Engineering Technology.

\subsection{Goals for Learners}
\begin{figure}[htbp]
\plotone{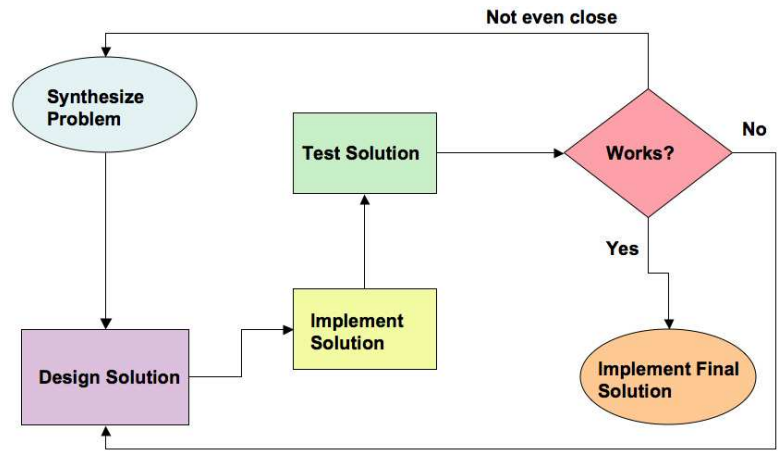}
        \caption{
                \label{fig:eng}
                The engineering problem-solving process.
                }
\end{figure}

As prior knowledge, earlier in the semester, students will have studied series and parallel circuits and Ohm's law,
and will have used multimeters and power supplies.
The content goals for this lab are understanding voltage dividers and Kirchoff's voltage law,
analyzing resistive networks (Th\'evenin equivalence),
and using the Wheatstone Bridge circuit to solve an engineering problem.
The process goals include building a circuit from a schematic diagram,
testing a circuit using multimeters, and utilizing the engineering problem-solving process
(particularly implementation, testing, and evaluation of a solution; see Figure~\ref{fig:eng}).
Attitudinal goals were teamwork, self-confidence in engineering skills, and motivation by real-world applications.

\subsection{Activity Description}

\begin{figure}[htbp]
\plottwo{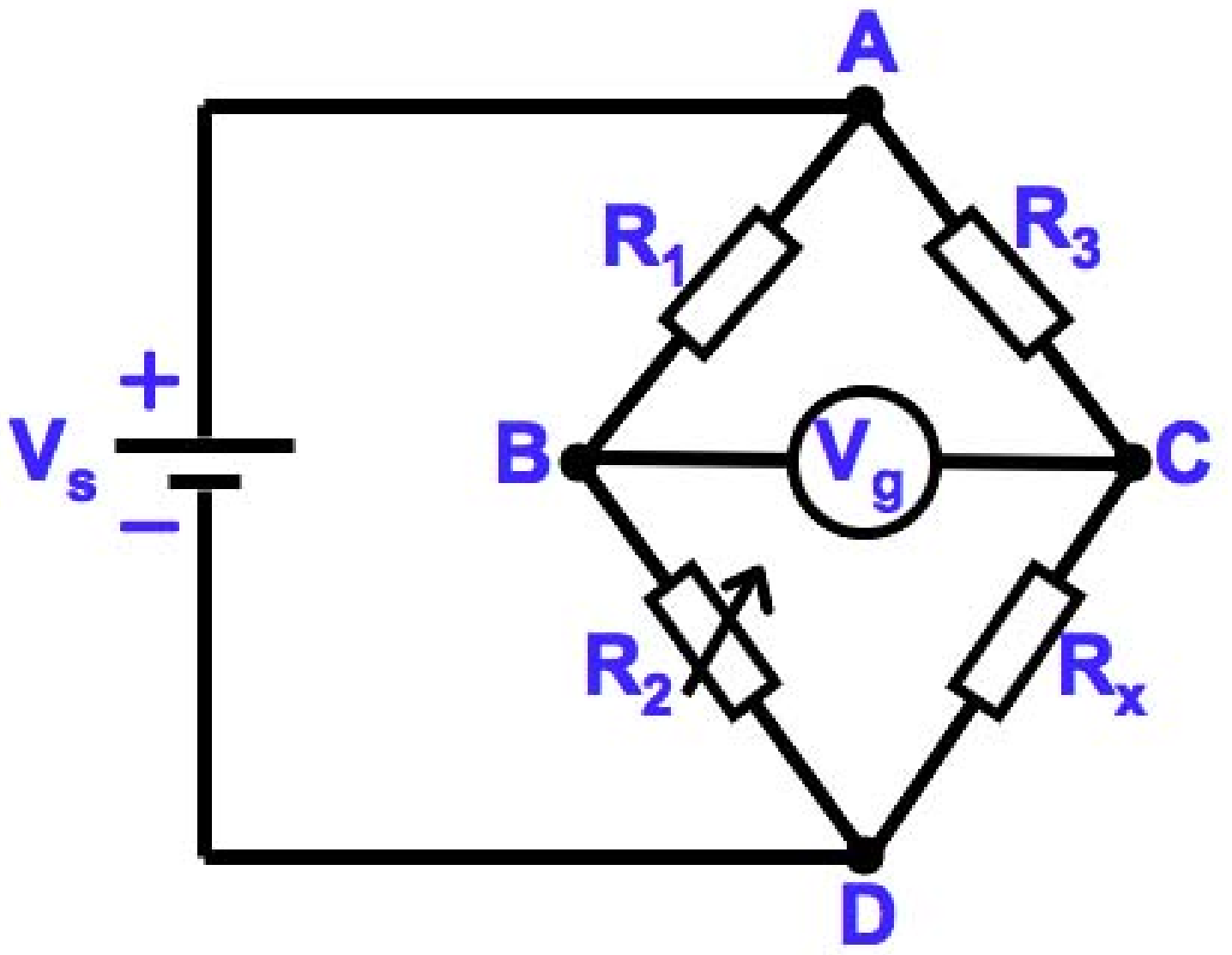}{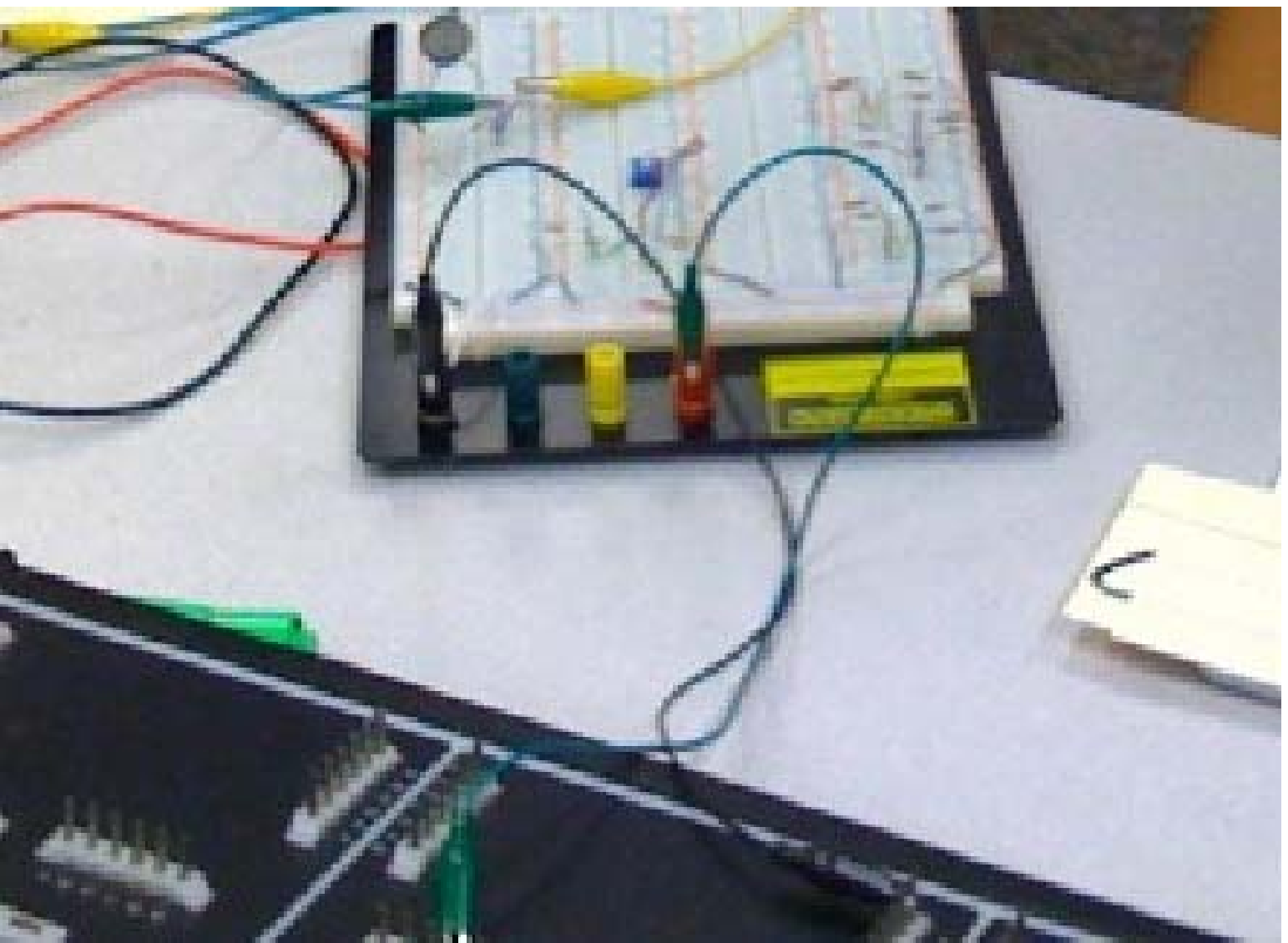}
        \caption{
                \label{fig:wheatstone}
                Wheatstone bridge circuit.  (Left) Schematic diagram.  (Right) Breadboard.
                }
\end{figure}

Figure~\ref{fig:wheatstone} shows the Wheatstone bridge circuit.
Four resistors ($R_1$, $R_2$, $R_3$, and $R_x$) are connected between four circuit junctions ($A$, $B$, $C$, and $D$).
Resistors $R_1$ and $R_2$ are connected in series;
resistors $R_3$ and $R_x$ are connected in series;
and the two legs are parallel to each other ($ABD \parallel ACD$).
The source voltage is $V_s$ and the output voltage $V_g$ is measured between $B$ and $C$.
When the voltage across the bridge is zero ($V_g = 0$), the bridge is said to be \textit{balanced}
and the ratios of resistances are equal:
\begin{equation}\frac{R_1}{R_2} = \frac{R_3}{R_x}.\end{equation}
In the specific example of Figure~\ref{fig:wheatstone}, $R_2$ is a variable resistor
(a rheostat or a potentiometer) and $R_x$ is a resistor of unknown value.
$R_x$ will also be referred to as $R_4$ for generalized Wheatstone bridges.

\subsubsection{Starters}

\begin{figure}[htbp]
\plottwo{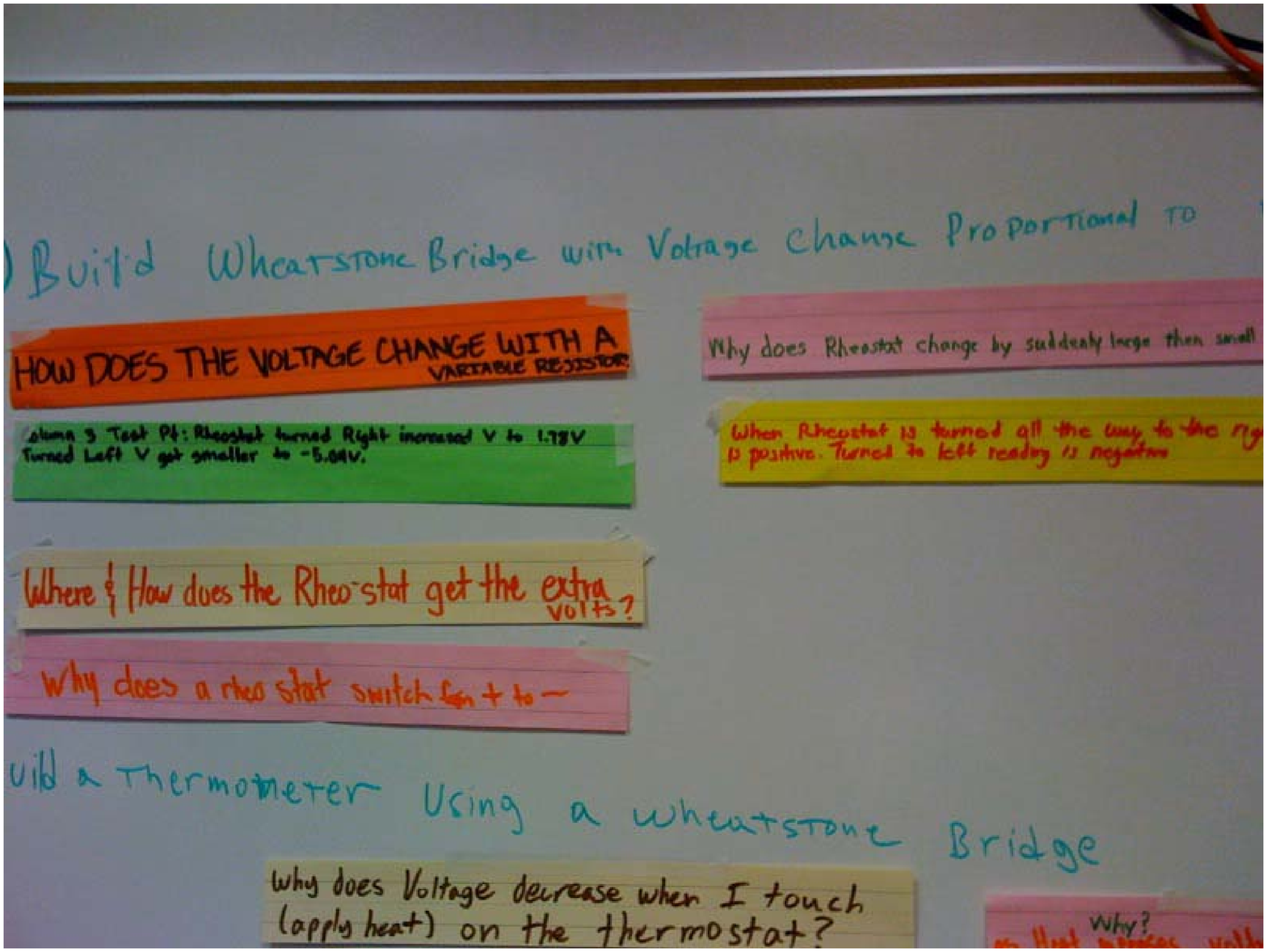}{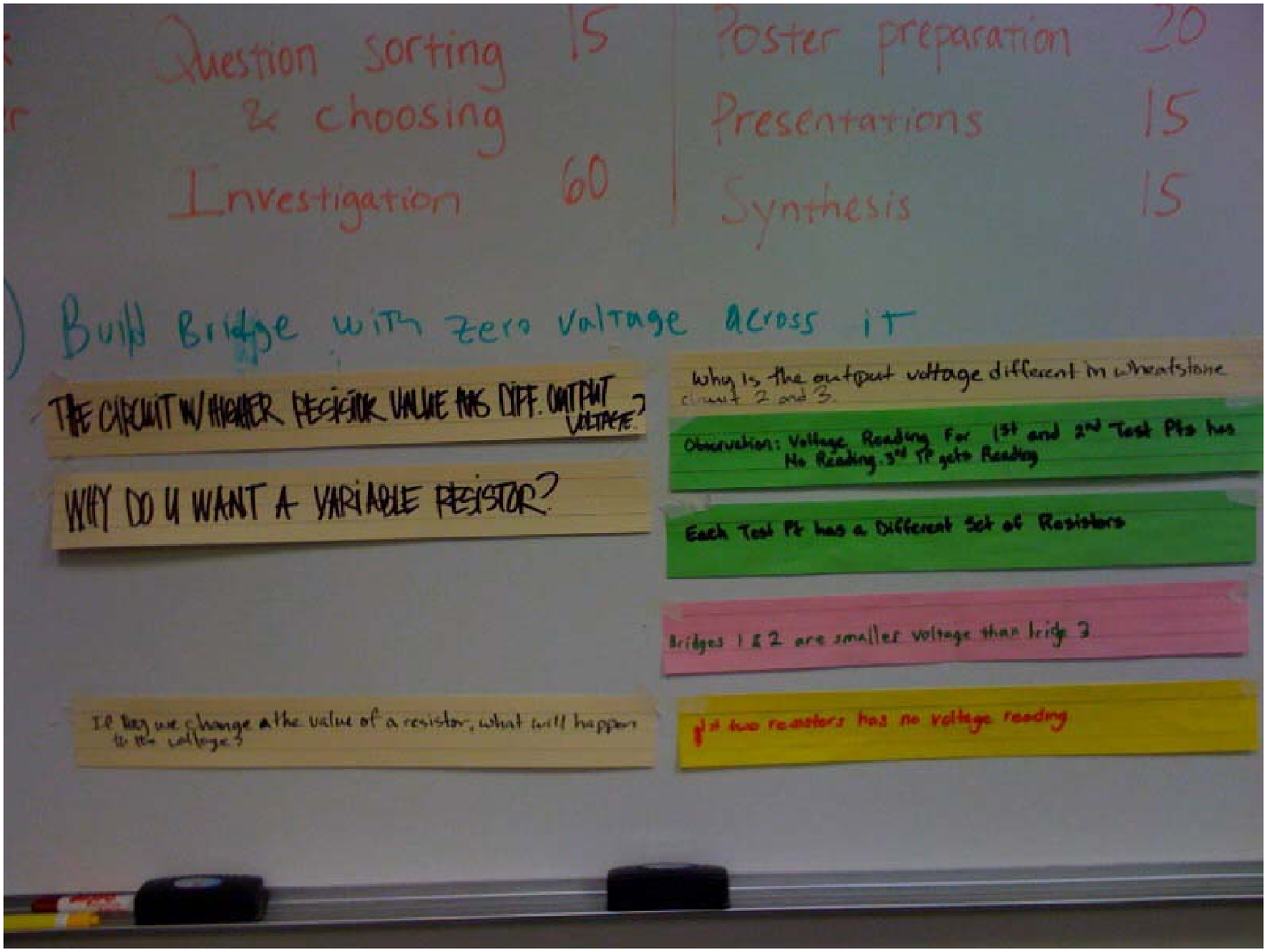}
        \caption{
                \label{fig:questions}
                Questions and observations generated by the students during the Starters.
                }
\end{figure}

The purpose of Starters in an inquiry is to generate curiosity and interest in students
by exposing them to new phenomena or content they will be working with in the Focused Investigation.
Facilitators (instructors) give directions about what to explore with each Starter,
and students are encouraged to write their observations and questions down on sentence strips (Figure \ref{fig:questions}).
In the Starters for this inquiry we introduce the students to the Wheatstone bridge,
both in terms of what it means when the circuit is balanced as well as its real-world applications.
First, a schematic is shown (Figure~\ref{fig:wheatstone}, left) during the introductory presentation.
Next, students see the circuit on a breadboard (Figure~\ref{fig:wheatstone}, right).
During the Starters, students observe and measure the bridge in a few different configurations,
with two students per station (i.e., two students per breadboard).
Table~\ref{tab:starters} describes the three Starters that are set up in three columns on the breadboard.

\begin{table}[htbp]
  \caption{
  Starters, column on breadboard, and Wheatstone circuit set-up.
  }
  \smallskip
  \centering
  \begin{tabular}{lcr}
    \hline
    \textbf{Starter Name} & \textbf{Column} & \textbf{Set-up of Circuits}\\
    \hline
    Balanced vs. Unbalanced & 1 & Two of $V_g=0$, one of $V_g \ne 0$\\
    Thermistor & 2 & $R_2$ is a thermistor\\
    Variable Resistor and Linearity & 3 & $R_2$ is a variable resistor\\
    \hline
  \end{tabular}
  \label{tab:starters}
\end{table}

In Starter 1 (Balanced vs.\ Unbalanced), three Wheatstone bridges are presented on a breadboard in column 1.
The first two bridges are balanced (output voltage $V_g = 0$) and the third bridge is unbalanced
(output voltage $V_g \ne 0$).
The two balanced bridges are balanced in different ways:
one is balanced with all resistors being equal ($R_1=R_2=R_3=R_4$),
while the other is balanced with different values of resistors in equal ratios
($R_1/R_2 = R_3/R_4$).
Students measure the output voltage and observe its dependence on resistance.
Note that when setting up the Starters, facilitators must be careful in choosing the appropriate resistors,
as small differences in resistance lead to easily noticeable changes in output voltage, as the bridge is very sensitive to resistance.

In Starter 2 (Thermistor) in column 2 of the breadboard, a thermistor (temperature-dependent resistor)
is placed at $R_2$ and students measure the output voltage as they warm up the thermistor with their hands.
Resistors $R_1$, $R_3$, and $R_x$ should be chosen in the linear regime such that students can balance the bridge
by varying $R_2$ at room temperature.

\begin{figure}[htbp]
\plotone{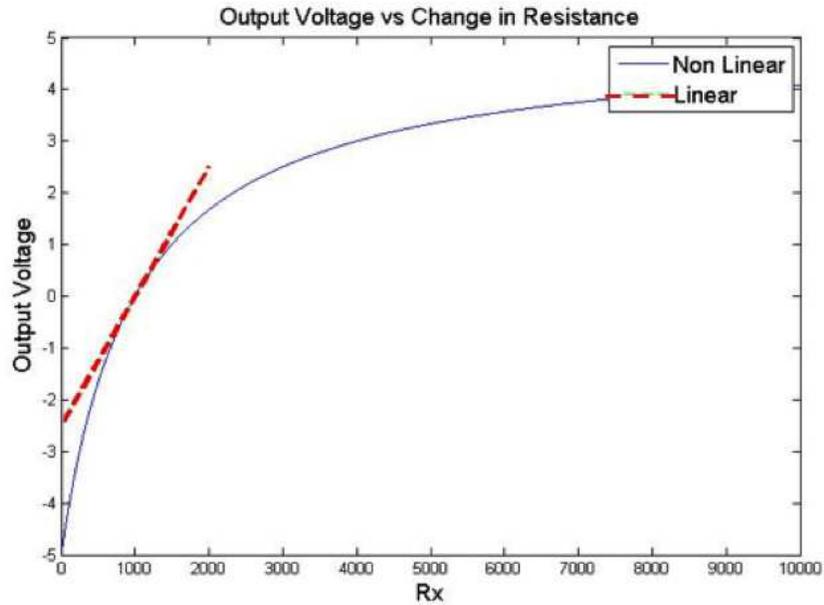}
        \caption{
                \label{fig:linearity}
                Output voltage $V_g$ as a function of variable resistance $R_x$.
                The Wheatstone bridge circuit is balanced where $V_g=0$
                and the linear regime is where $V_g$ is approximately proportional to $R_x$.
                }
\end{figure}

Starter 3 (Variable Resistor and Linearity) is set up in column 3 of the breadboard,
with a rheostat or potentiometer used at $R_2$.
Students again measure the output voltage, this time observing the effect while they vary the resistance $R_2$.
The linearity of output voltage (Figure \ref{fig:linearity}) with resistance can be explored with this circuit.

Students write observations and questions down on sentence strips while doing the Starters.
Facilitators help with the setup and with using the multimeter to measure output voltage at $V_g$.
During a short break, facilitators sort the questions into the categories under the Engineering Challenges in Table~\ref{tab:challenges}.
Sample questions sorted into the categories are shown in Table~\ref{tab:questions}.

\subsubsection{Focused Investigation}

Table~\ref{tab:challenges} lists the engineering challenges offered as options for the students' investigations.

\begin{table}[htbp]
  \caption{
  Engineering challenges with a Wheatstone bridge.
  }
  \smallskip
  \centering
  \begin{tabular}{l}
    \hline
    \textbf{Challenges for investigation}\\
    \hline
    A.  Build a perfectly balanced Wheatstone bridge\\
    B.  Build a Wheatstone bridge for operation in a linear regime\\
    C.  Use a Wheatstone bridge to build a thermometer\\
    D.  Use a Wheatstone bridge to determine an unknown resistance\\
    \hline
  \end{tabular}
  \label{tab:challenges}
\end{table}

\begin{table}[htbp]
  \caption{
  A partial list of the questions and observations generated during the Starters by students,
  each categorized into one of four design goals A-D.
  }
  \smallskip
  \centering
  \begin{tabular}{l}
    \hline
    \textbf{Build a perfectly balanced Wheatstone bridge}\\
    First two circuits in Starter 1 have no voltage reading\\
    Why is the output voltage different in circuit 2 and 3?\\
    Each test point has a different set of resistors\\
    The circuit with higher resistor value has diff. output voltage?\\
    \hline
    \textbf{Build a Wheatstone bridge with voltage change proportional to resistance}\\
    How does the voltage change with a variable resistor?\\
    Why does rheostat change by suddenly large then small amounts?\\
    Why does a rheostat switch from + to - ?\\
    \hline
    \textbf{Use a Wheatstone bridge to build a thermometer}\\
    Why does voltage decrease when I touch (apply heat) on the thermistor?\\
    How does temperature affect a thermistor circuit?\\
    \hline
    \textbf{Use a Wheatstone bridge to build an Ohmmeter}\\
    If we change the value of a resistor, what will happen to the voltage?\\
    Why do u want a variable resistor?\\
    Purpose of the Wheatstone bridge?\\
    \hline
  \end{tabular}
  \label{tab:questions}
\end{table}

The focused investigation is the heart of the inquiry and goes on for the rest of Day 1 and the first half of Day 2.
After student teams have chosen a design goal from Table 2, they work to achieve that goal using the breadboards and circuitry available.
(Students may not re-use the Starter circuits but must build their own circuits from scratch,
since building a circuit from schematic diagram is an important process skill in this lab.)

The differences between the four engineering challenges in Table \ref{tab:challenges} are choice of resistors.
All students will build Wheatstone bridges, but the specifics of $R_1$, $R_2$, $R_3$, and $R_4$ will vary.
During much of the activity, students will test the output voltage at $V_g$
and then swap or adjust their resistors to get the desired value.
Students may need heavy facilitation during the building stage if they are unfamiliar with breadboard circuitry or need help using multimeters for measurements.

In design goal A, ``Build a perfectly balanced Wheatstone bridge,''
the output voltage $V_g$ of the bridge circuit should read exactly zero.
This is achieved by carefully measuring and inserting resistors with the exact ratios $R_1/R_2 = R_3/R_4$.

In design goal B, ``Build a Wheatstone bridge for operation in a linear regime,''
a variable resistor (a rheostat or potentiometer) should be used for $R_2$,
and the other three resistors should be selected to ensure the output voltage is in the linear regime
for a good fraction of the range of the variable resistor (Figure \ref{fig:linearity}).
The resulting Wheatstone bridge should have a voltage change proportional to resistance.
The slope of the proportionality depends on the choice of resistors, and
so it should be linear over a large variation of the rheostat.

In design goal C, ``Use a Wheatstone bridge to build a thermometer,''
a thermistor is used as resistor $R_2$,
and the output voltage $V_g$ can determine temperature once the thermistor is calibrated (a conventional thermometer must be used to calibrate the Wheatstone thermometer).
Ice packs wrapped in absorbent cloth (to avoid condensation shorting the circuit)
and a hair dryer or the outside of a coffee cup can be used to provide temperature variation.

In design goal D, ``Use a Wheatstone bridge to determine an unknown resistance,''
the bridge is used in an unbalanced condition.
Three resistors are known ($R_1$, $R_2$, and $R_3$)
and any unknown resistor can be inserted into the fourth position ($R_x$). The output voltage will vary, and this change can be used to determine the unknown resistance.

At the beginning of the second day, we presented a thinking tool: a short lecture about Ohm's law ($V=iR$)
and Kirchoff's laws (voltage law and current law) so the students can
calculate the output voltage
if the values of the resistors are known.

\subsubsection{Sharing and Synthesis}

Students conclude their investigations by making posters to present what they learned with the rest of the class.
Column 1 in Table \ref{tab:sharing} lists the requirements for the poster presentations,
and was written on the board for the students.
Column 2 in Table \ref{tab:sharing} shows the correspondence of the poster requirements
with the rubric we used to assess presentations.

\begin{table}[htbp]
  \caption{
  Final poster requirements for sharing, and correspondence to categories assessed with the rubric.
  }
  \smallskip
  \centering
  \begin{tabular}{clr}
    \hline
    \textbf{Item} & \textbf{Poster} & \textbf{Correspondence}\\
    \textbf{\#} & \textbf{Requirement} & \textbf{to Rubric}\\
    \hline
    1 & State your design goal. &  (none)\\
    \hline
    2 & Draw schematic of your circuit such that &  Support\\
       &  someone else could build it.  & \\
    \hline
    3 & State why your circuit meets your design  & Solution\\
     & goal.  & \\
    \hline
    4 & Explain what was important about choosing  &  Reasoning/\\
      &   your resistors ($R_1$, $R_2$, $R_3$, and $R_4$) so that & Justification \\
        & your circuit meets your design goal.  & \\
    \hline
  \end{tabular}
  \label{tab:sharing}
\end{table}

During the synthesis, instructors tie the investigations together by clarifying the details of how to balance a Wheatstone bridge and its applications.
Each teams' work is referenced to point out how everyone learned something.

\section{Elements of Inquiry Design for Engineering}

This activity illustrates differences in designing inquiry for engineering as opposed to designing inquiry for science.
        
\subsection{Starters: Questions become Design Goals}

In the Starters for the quintessential Light and Shadows inquiry usually studied by first-year PDP participants,
learners are shown various phenomena related to light and shadows,
and write their questions on sentence strips.
During the gallery phase, learners then choose one of the questions to investigate.
For the Circuit Design inquiry, learners wrote questions and observations on sentence strips as in Light and Shadows.
However, these questions were then sorted into the four categories listed in Table \ref{tab:challenges}
as Engineering Challenges or Design Goals.
We utilized a different format more appropriate to engineering, as engineering is more the application of scientific principles.
Therefore, our modification of the use of questions generated during the starters is an authentic
adjustment for an engineering inquiry.

\subsection{The Rubric as a Tool for Assessment}

\subsubsection{Claim, Evidence, Reasoning becomes Solution, Support, Reasoning}

We also modified the evaluation rubric for the engineering inquiry. The rubric PDP participants used to assess students' learning in a science inquiry is designed to evaluate how students answered their question; the categories for evaluation are claim, evidence, and reasoning.

An engineering activity is more concerned with how a student was able to solve their problem and accomplished their design challenge; so the categories became proposed solution, support (including tradeoffs and optimization), and reasoning or justification of how the solution worked.

For the ``support'' category in this case we decided to focus on the students' understanding
of the Wheatstone bridge by drawing a schematic diagram of their circuit
(similar to Figure \ref{fig:wheatstone} left)
or demonstrating understanding of the equation for calculating the output voltage of their circuit:
\begin{equation}
V_g = V_s \left ( \frac{R_x}{R_x+R_3} - \frac{R_2}{R_1+R_2} \right ).
\end{equation}
For the ``Solution'' we were looking for a statement of how students met their design goal.
For the ``Reasoning'' we were looking for an explanation of the balance of resistance
and how the output voltage $V_g$ depends on the resistor values $R_1$, $R_2$, $R_3$, and $R_x$.
The rubric is shown in Table~\ref{tab:rubric}.

\begin{table}[htbp]
  \caption{
  Rubric used for summative assessment.
  }
  \small
  \smallskip
  \centering
  \begin{tabular}{m{0.01cm}m{1.4cm}m{2.2cm}m{2.2cm}m{2.2cm}m{2.2cm}m{0.01cm}}
    \hline
    & & \textbf{Off-track} & \textbf{Emerging} & \textbf{Accomplishing} & \textbf{Mastering} & \\
    & & \textbf{[0]} & \textbf{[1]} & \textbf{[2]} & \textbf{[3]} & \\
    \hline
    & \raggedright{\textbf{Solution (Claim)}} & \raggedright{Solution does not address their question} & \raggedright{Solution not correct OR Solution may work but is convoluted (i.e., redundancy)} & \raggedright{Solution works / is correct} & \raggedright{Solution works / is correct AND state application, or state range of validity and linearity} & \\
    \hline
    & \raggedright{\textbf{Support / Circuit analysis}} & \raggedright{No diagram or equation} & \raggedright{Diagram with some errors} & \raggedright{Diagram OR equation} & \raggedright{Diagram and equation} & \\
    \hline
    & \raggedright{\textbf{Reason\-ing / Justification}} & \raggedright{Not investigating balanced vs.\ unbalanced or output voltage as a function of the 4 resistance legs} & \raggedright{Incomplete understanding of $V_g$ as a function of $R_1$, $R_2$, $R_3$, and $R_x$} & \raggedright{Understanding of interaction of components and mechanism of Wheatstone bridge (balanced vs. unbalanced)} & \raggedright{Meets ``accomplishing'' PLUS efficiency of design; or Tradeoffs; or Linear regime} & \\
    \hline
  \end{tabular}
  \label{tab:rubric}
\end{table}

\normalsize

\subsubsection{Reliability and Validity}

Scoring using the rubric was challenging,
but it was important for discerning whether it was a reliable and valid test of students' learning.
A valid test accurately characterizes what the students learned, while a reliable test gives similar scoring across time and assessor variation.

Table~\ref{tab:results} shows the scores given to each of the six teams
by each of the six scorers (facilitators and teaching consultant).
The maximum score possible was nine (9) points, and the mean class score was 6.2 points.
The standard deviation provides some measure of reliability.
The mean class standard deviation was 1.2 points, or 13\% out of 9 points.
This is more than one point uncertainty, implying that the rubric may not have been a reliable test and can be improved.

\begin{table}[htbp]
  \caption{
  Summative assessment results.  Each team of students was scored by each facilitator using the rubric.
  }
  \smallskip
  \centering
  \begin{tabular}{ccccccc}
    \hline
    Scorer & Team 1 & Team 2 & Team 3 & Team 4 & Team 5 & Team 6\\
    \hline
    A & 7 & 8 & 6 & 6 & 7 & 6\\
    B & 9 & 6 & 7 & 7 & 7 & 4\\
    C & 8 & 8 & 6.5 & 4 & 5 & 6\\
    D & 8 & 4.5 & 6 & 5.5 & 5.5 & 2.5\\
    E & 6 & 6 & 5 & 7 & 4 & 5\\
    F & 6 & 8 & 7 & 7 & 6 & 5\\
    \hline
    \textit{Mean} & \textit{7.3} & \textit{6.8} & \textit{6.3} & \textit{6.1} & \textit{5.8} & \textit{4.8} \\
    \textit{Std. Dev.} & \textit{1.2} & \textit{1.5} & \textit{0.8} & \textit{1.2} & \textit{1.2} & \textit{1.3}\\
    \hline
  \end{tabular}
  \label{tab:results}
\end{table}

\section{Conclusions}

The Circuit Design inquiry featured students learning important engineering process skills: they built a circuit on a breadboard to implement and test an electrical engineering question.
Our modifications to adjust from a science inquiry to an engineering inquiry included organizing student-generated questions under Design Goal challenge-style categories, and modifying the rubric to emphasize solving problems.
While it may not have provided reliable scores, the establishment of an engineering rubric helped immensely in clarifying our activity design goals and facilitation emphases.
Overall, the inquiry went well and accomplished many of the learning goals, and the students seemed to enjoy it.
Labs such as these, inserted into formal courses at UH-Maui,
are helping to train future STEM workers for the tech industry on the island of Maui.

\begin{figure}[htbp]
\plotone{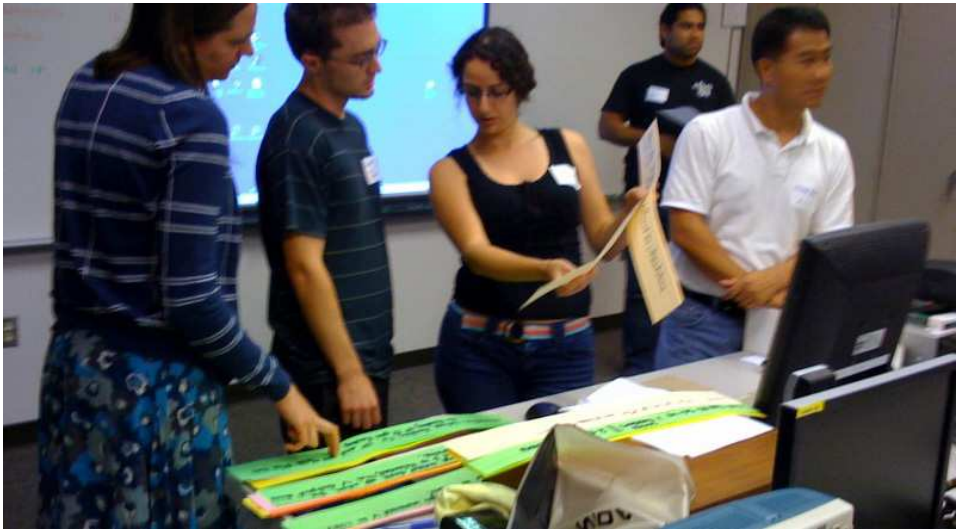}
        \caption{
                \label{fig:facilitators}
                Facilitators KM, CD, TF, OA, and JP sorting students' questions.
                }
\end{figure}

\acknowledgements
Design team member Vivian U was not able to travel to Maui to teach the lab, but she helped
as much as the other team members (Figure \ref{fig:facilitators}) in designing the inquiry.
Lisa Hunter was an observer and design-team consultant while at MCC,
 while Patrik Jonsson was a consultant at the PDP workshop.
 UH-Maui professors Mark Hoffman (the classroom teacher hosting this inquiry)
 and Elisabeth Reader provided advice and assistance.  This material is based upon work supported by: the National Science Foundation (NSF) Science and Technology Center program through the Center for Adaptive Optics, managed by the University of California at Santa Cruz (UCSC) under cooperative agreement AST\#9876783; NSF AST\#0836053; NSF AST\#0850532; NSF AST\#0710699; Air Force Office of Scientific Research (via NSF AST\#0710699); UCSC Institute for Scientist \& Engineer Educators; and the University of Hawai`i.

\bibliography{morzinski}

\end{document}